\pdfoutput=1
\documentclass{JINST}
\usepackage{graphicx}

\title{High granularity tracker based on a Triple-GEM 
optically read by a CMOS-based camera}

\author{M. Marafini$^{a,b}$, 
V. Patera$^{a,b,d}$, 
D. Pinci$^a$\thanks{Corresponding author.}, 
A. Sarti$^{b,c,d}$, 
A. Sciubba$^{a,b,d}$
and E. Spiriti$^{c}$\\
\llap{$^a$}Istituto Nazionale di Fisica Nucleare\\ 
Sezione di Roma, I-00185, Italy\\
\llap{$^b$}Museo Storico della Fisica e Centro Studi e Ricerche "Enrico Fermi" \\
Piazza del Viminale 1, Roma, I-00184, Italy\\
\llap{$^c$}Istituto Nazionale di Fisica Nucleare \\ 
Laboratori Nazionali di Frascati, I-00040, Italy\\
\llap{$^d$}Dipartimento di Scienze di Base e Applicate per l'Ingegneria\\ 
Sapienza Universit\`a di Roma, I-00185, Italy\\

E-mail: \email{davide.pinci@roma1.infn.it}}

\abstract{
The detection of photons produced during the avalanche
development in gas chambers has been the subject
of detailed studies in the past.
The great progresses achieved in last years
in the performance of micro-pattern gas detectors on one side
and of photo-sensors on the other
provide the possibility of making high granularity 
and very sensitive particle trackers.

In this paper, the results obtained with 
a triple-GEM structure read-out
by a CMOS based sensor are described. 
The use of an He/CF$_4$ (60/40) gas mixture
and a detailed optimization of the electric 
fields made possible to obtain a very high 
GEM light yield.
About 80 photons per primary electron were detected by 
the sensor resulting in a very good capability of tracking
both muons from cosmic rays and electrons from natural radioactivity.}


\begin{document}

\section*{Introduction}

Micro-pattern gas detectors have proven to be versatile devices for
high resolution particle tracking. One of the most successful
micro-pattern technologies developed so far
is the Gas Electron Multiplier (GEM),
introduced in 1996 at CERN \cite{Sauli:1997qp}. 
Electrons produced in a gas mixture 
by ionizing particles are
multiplied within the GEM foil channels 
where a high
electric field is present.
During the multiplication process, 
photons are produced along with electrons 
by the gas through atomic and molecular de-exitation.

CMOS based photo-sensors can be exploited to read-out this light.
Given their impressive development in last years,
commercial CMOS detectors offer a very high granularity 
together with a very low noise level.
Primary electrons, produced by ionizing particles,
are multiplied in a GEM based structure and 
the image of the holes where the avalanches happen
can be acquired. In this way, not only the total amount 
of light (proportional to the energy realeased in gas),
but also the position of the avalanches can be recorded.
The projection of the track on the GEM plane
can be acquired and used to reconstruct the particle
path inside the detector.

\section{Experimental set-up}
\subsection{The triple-GEM structure}
\begin{figure}[!h]
\centering
\includegraphics[width=0.89\linewidth]{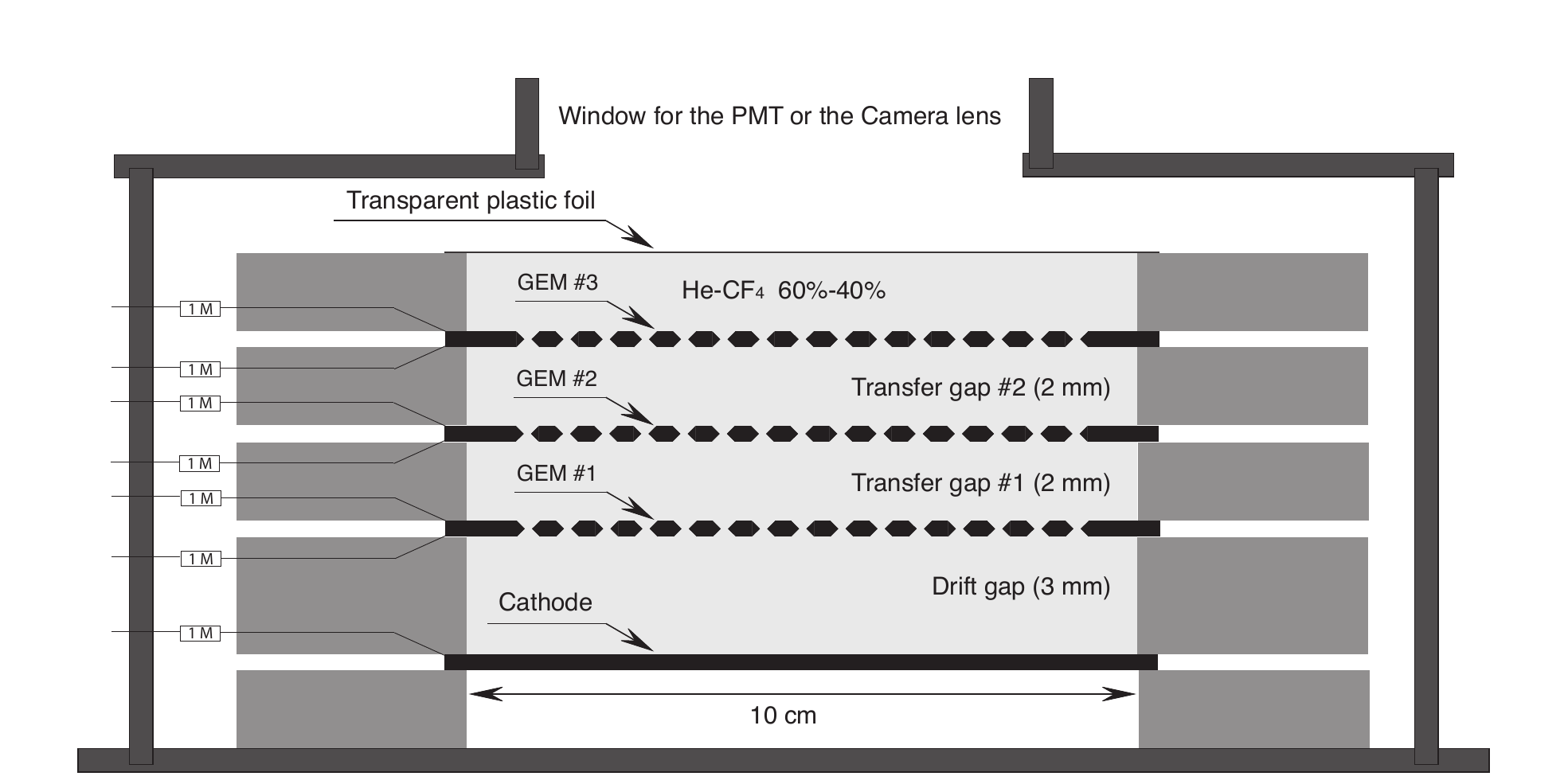}
\caption{Drawing (not to scale) of the triple-GEM stack.
The drift and transfer gaps are shown.}
\label{fig:setup}
\end{figure}

The electron multiplication structure was 
obtained by stacking three 10$\times$10 cm$^2$
standard GEM foils 
(70 $\mu$m diameter holes with 140 $\mu$m pitch)
one above the other as shown in
Fig. \ref{fig:setup}.

A planar cathode was placed in front of
the first GEM to create a 3~mm wide drift gap
that represents the main sensitive volume of the system.
Two 2~mm wide transfer gaps
were left between the three GEM foils.
The readout plane was replaced by a transparent 
plastic foil in order to allow a photo-detection
from outside.
No electric field was foreseen between the 
last GEM and the plastic foil.
The electrons created in the multiplication 
were collected on the electrodes of the third GEM. 
Seven different high voltage channels were used to supply
the detector through 1 M$\Omega$ protection resistors.
The GEM structure was contained in a black box equipped
with an upper window used for the image acquisition.

\subsection{The gas mixture}
\label{sec:gas}
\begin{figure}[!h]
\centering
\includegraphics[width=0.49\linewidth]{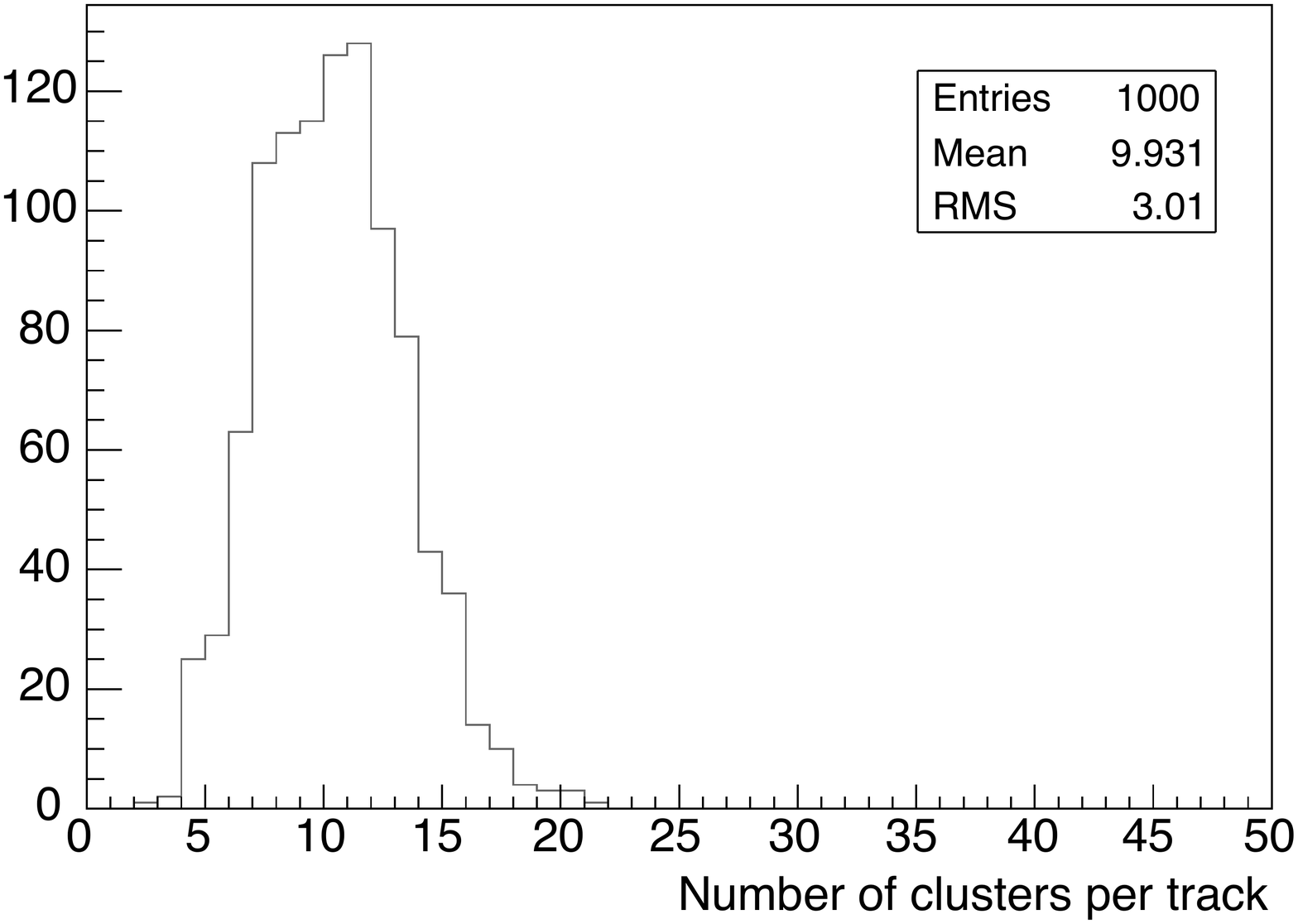}
\includegraphics[width=0.49\linewidth]{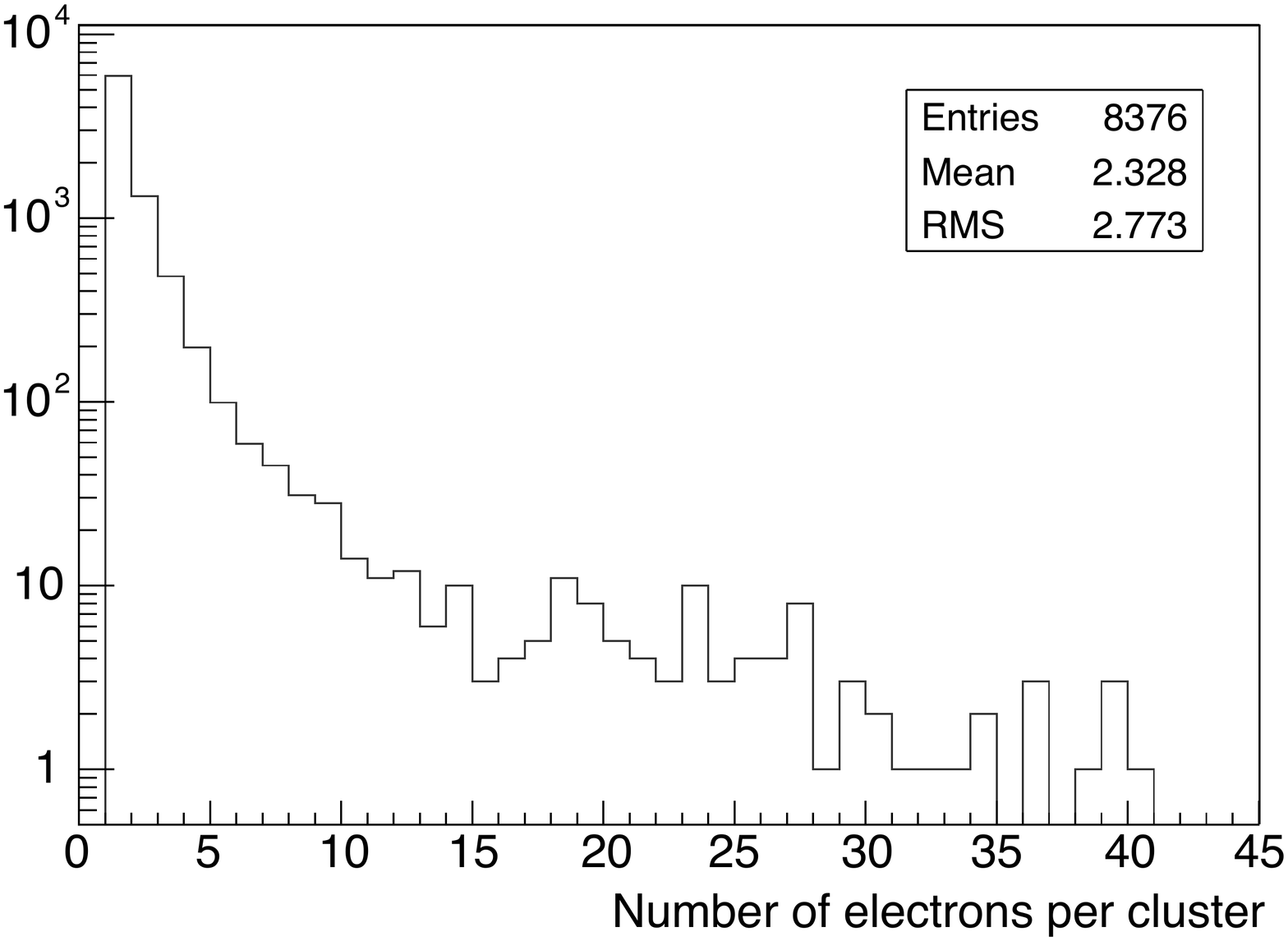}
\caption{Results of the Garfield simulation:
distribution of number of ionization clusters created by
2 GeV muons in the 3~mm wide drift gap (left) and 
distribution of the number of electrons per cluster (right).} 
\label{fig:gas}
\end{figure}

All measurements presented in this paper
were performed by using a binary gas mixture 
He/CF$_4$ (60/40). This mixture is expected to have an
emission spectrum with a main contribution around 600 nm
characteristic of the CF$_4$ \cite{bib:fraga}.
In order to evaluate the
properties of the gas mixture, 
a Garfield simulation  \cite{bib:garfield}
was performed.
The ionization processe due to
the crossing of 2 GeV muons in the 3~mm drift gap
was studied.
As shown in Fig.~\ref{fig:gas} (left) these muons
produce about 10 clusters in average in the 3~mm drift gap.
Therefore, the mean distance between two ionization 
points is about 300 $\mu$m.

The distribution of the number of electrons per cluster 
(Fig. \ref{fig:gas}, right) has a mean
value of 2.3. 
The total number of primary electrons due to a
minimum ionizing muon crossing the drift gap
perpendicularly to the GEM plane is 
thus expected to be around 20.

\section{Light yield optimization and measurements}

\subsection{Signal acquisition and analysis}
In order to evaluate the amount 
of light produced by the GEM stack, 
a first set of measurements was
performed by acquiring the light
with an R9800 photo-multiplier\footnote{www.hamamatsu.com/us/en/R9800.html}
with a 25 mm window (GEM-PMT in the following).
The detector was placed with the GEM foils in horizontal position
(as shown in Fig. \ref{fig:setup}) so that vertical muons travel
about 3 mm in the drift gap.
The system was tested by using cosmic rays externally triggered
by two NaI scintillators, 
placed one above and one below the GEM structure.
The scintillators have a squared shape
with a 10 cm side.

\begin{figure}[!h]
\centering
\includegraphics[width=0.89\linewidth]{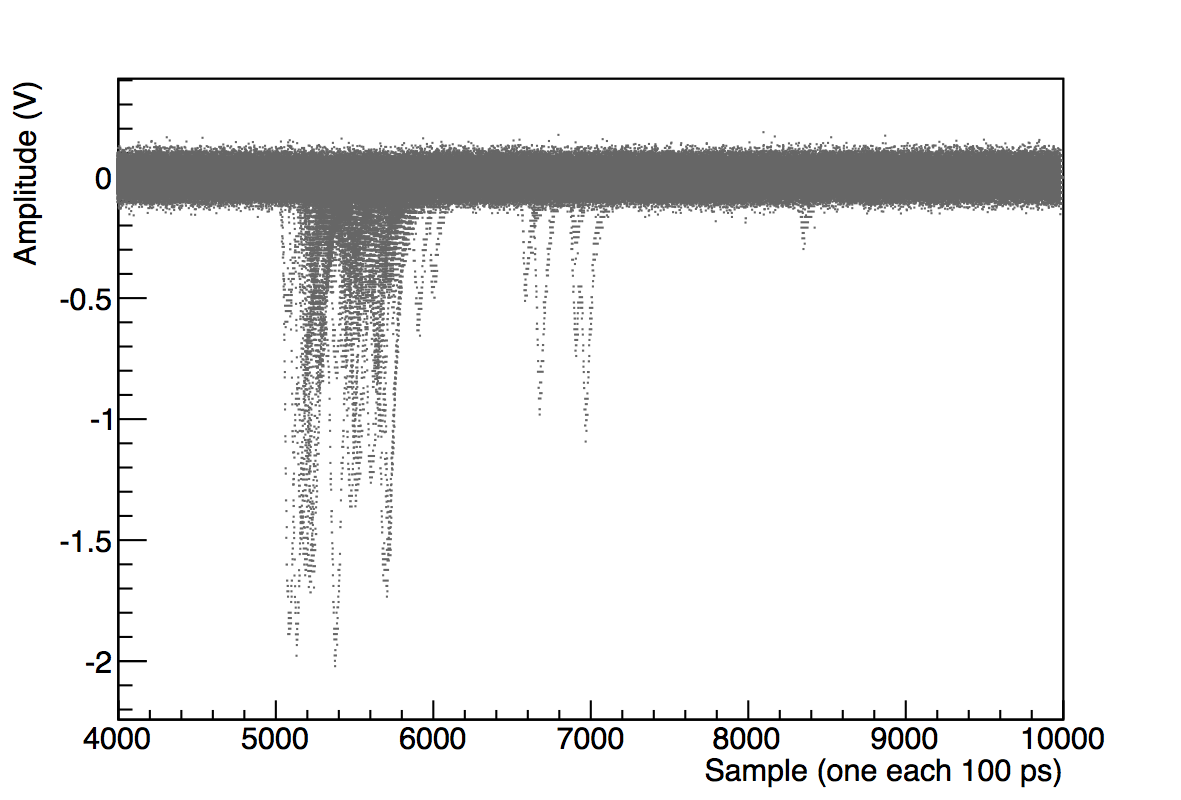}
\caption{Example of several superimposed waveforms
acquired from the GEM-PMT.}
\label{fig:scope}
\end{figure}

The waveforms of the signals provided by the two PMT
reading the NaI and the GEM-PMT 
were acquired by a Lecroy WavePro 7300
10 GS/s oscilloscope\footnote{teledynlecroy.com/search/default.aspx?q=7300}.
Figure \ref{fig:scope} shows several waveforms
of signals produced by the GEM-PMT.

\begin{figure}[!h]
\centering
\includegraphics[width=0.59\linewidth, angle=90]{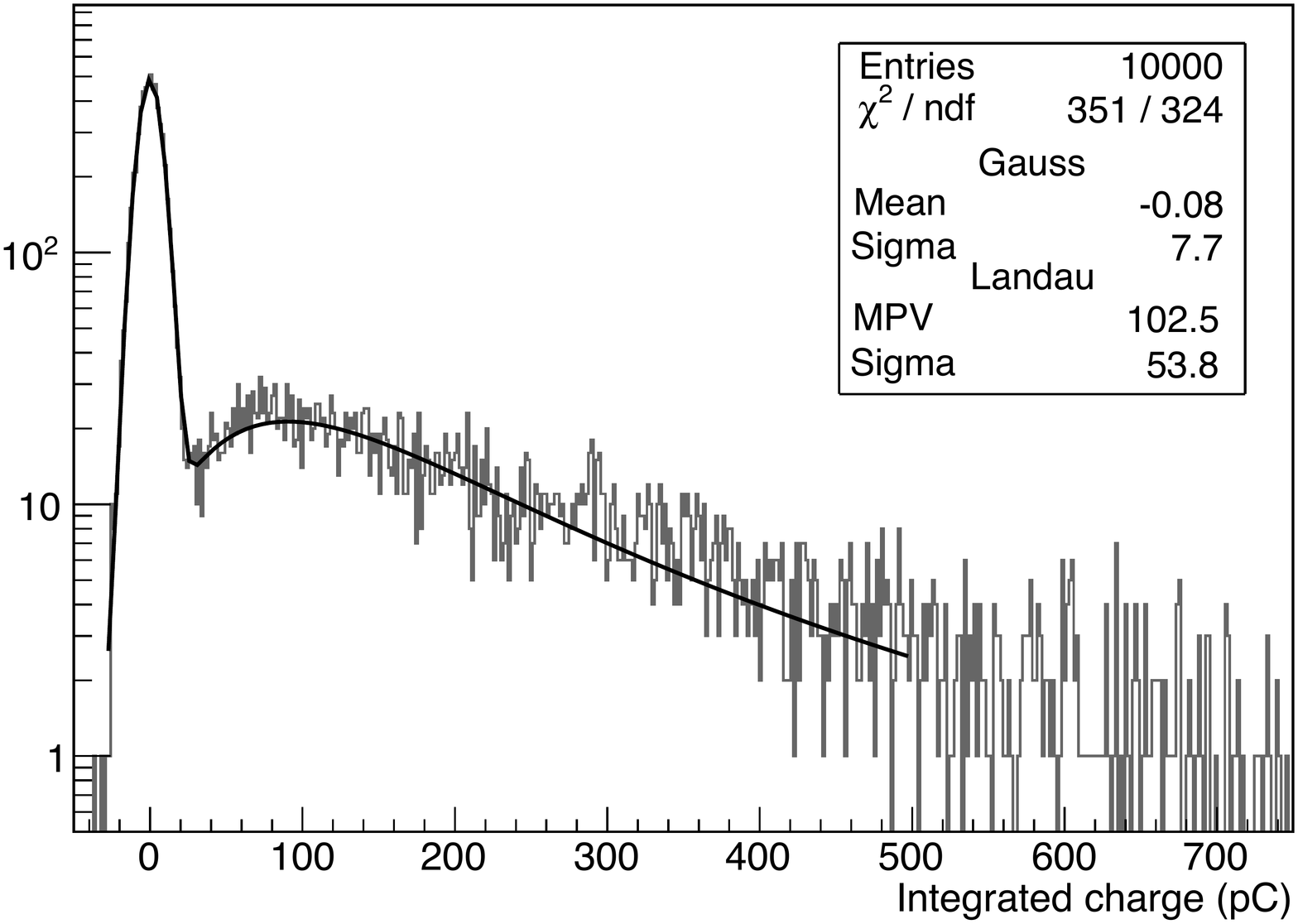}
\caption{Example of a charge spectrum obtained by integrating
the PMT waveforms with a superimposed "Gauss+Landau" fit.}
\label{fig:charge}
\end{figure}

Each trigger, a 1 $\mu$s wide time window 
(i.e. 10,000 samples) was acquired.
Trigger signal is synchronized with the center of the
time window. The first half of the window
(5,000 samples) was used to evaluate the noise
behavior while signals due to crossing muons
arrive in the second half of the window.
Signals between sample 5,000 and 6,000 are due to light 
produced by the GEM stack. 
Few late signals, due to secondary ionization within the detector
are also visible.
The acquired waveforms have been analysed to evaluate
the total charge provided by the GEM-PMT. 
Since the area of the trigger scintillators 
is larger than the GEM-PMT window 
in the main part of triggered events
light is produced far away from the GEM-PMT.
In Fig. \ref{fig:charge}
an example of a charge spectrum 
of the GEM-PMT
is shown, with a superimposed
"Gauss+Landau" fit.
The mean value of the Gauss curve gives the 
pedestal level while the most probable value (MPV) of the Landau curve 
is used to evaluate the charge (i.e. the light)
collected in each configuration.

\subsection{Optimization of electric fields}

The study of the behavior of the total charge
integrated in the PMT signals,
made it possible to optimize the
electric fields between the GEM 
in order to maximise the light yield.
The voltages across the three GEM foils
were kept at 480 V in all measurements.

\begin{figure}[!h]
\centering
\includegraphics[width=0.49\linewidth]{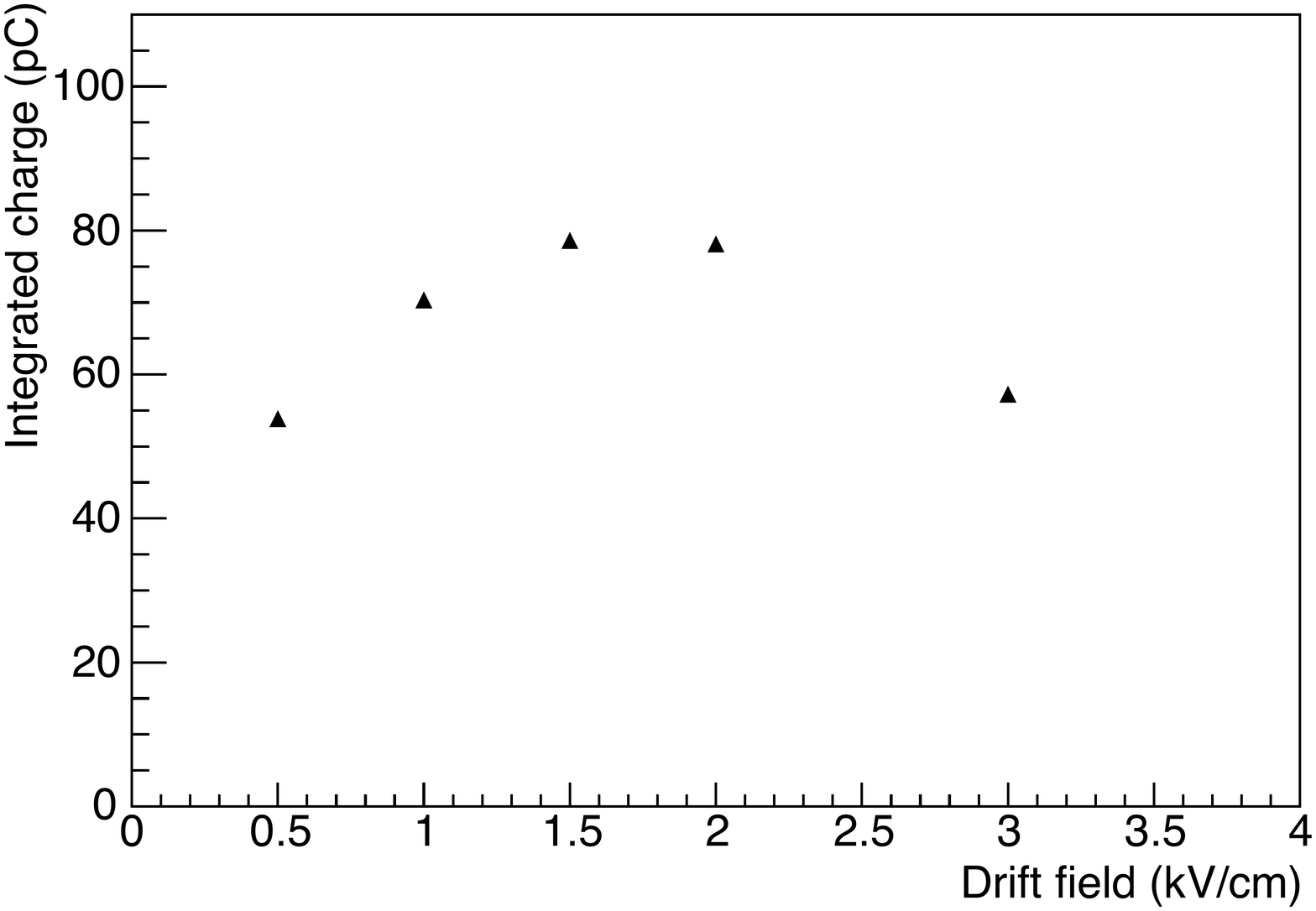}
\includegraphics[width=0.49\linewidth]{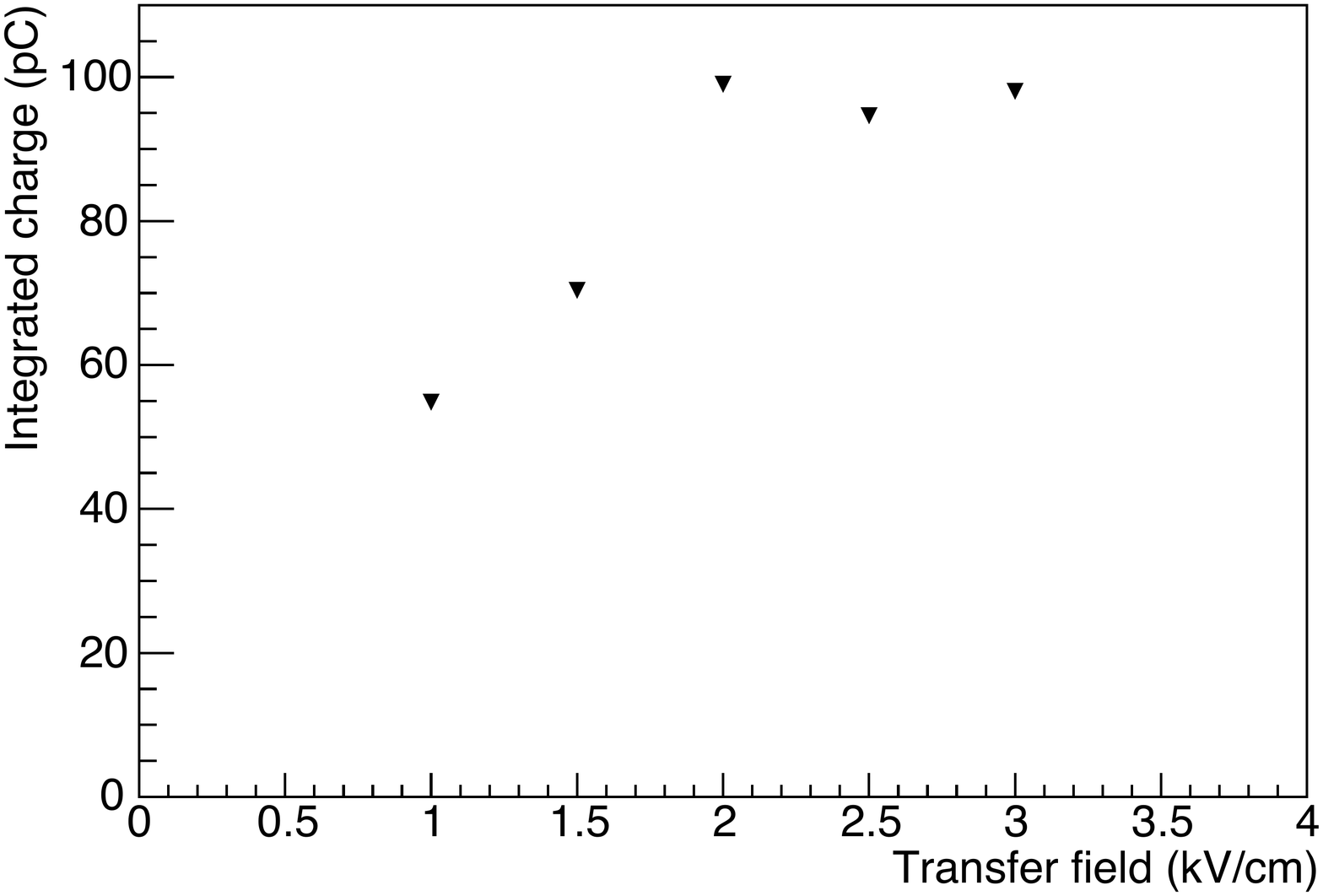}
\caption{Light yield as a function of the drift field (left)
and of transfer fields (right). Data uncertainties are
smaller than the markers.}
\label{fig:ly_E}
\end{figure}

In a first set of measurements,
the amount of collected light was recorded
by varying the drift field while keeping
the two transfer fields at a value of 1.5 kV/cm.
The result, on the left in Fig.~\ref{fig:ly_E},
shows a maximum for drift field values
between 1.5 kV/cm and 2.0 kV/cm.
For higher values
the light yield decreases
because a strong drift field reduces the
capability of the GEM in collecting electrons
within the multiplication holes (defocusing effect \cite{bib:tesi}).

The light yield dependence on the transfer fields
was then studied while keeping the drift field at 1.0 kV/cm
and by varying both the transfer fields. 
As shown on the right of Fig.~\ref{fig:ly_E},
the light yield increases very rapidly while increasing
the transfer fields and it is almost stable for values
in the range $2.0 \div 3.0$ kV/cm.
The charge spectrum in Fig.~\ref{fig:charge} was
obtained in the optimized conditions: 
drift field at 1.5 kV/cm and
transfer fields at 2.0 kV/cm.
The PMT response to a single photo-electron was
measured by using a calibrated light source
and was found to be $0.04 \pm 0.01$ pC.
Therefore it was calculated
that in the optimised field configuration,
as shown in Fig. \ref{fig:charge},
about 2500 p.e. were collected in average per
cosmic ray crossing.

\section{Measurements with a CMOS-based camera}

After the light yield was optimised, the acquisition of
images by means of a Hamamatsu CMOS-based 
camera\footnote{Orca flash 4.0. For more details visit www.hamamatsu.com} was performed.
This camera provides several qualities that make it an optimal choice
for this kind of applications:

\begin{itemize}
\item {\it low noise}: nominal level lesser than 2 photons per pixel;
\item {\it high sensitivity}: a quantum efficiency higher than 70\% in the 
CF$_4$ emission spectral range;
\item {\it high granularity}: 2048 $\times$ 2048 pixels  
with an area of 6.5 $\mu$m $\times$ 6.5$\mu$m for a total
sensitive surface of 13.3 mm $\times$ 13.3 mm. 
\end{itemize}

\subsection{Sensor performance studies}
\label{sec:pre}

The performance of the CMOS sensor have been investigated by
means of a calibrated light source.
A 1 mm diameter calibrated light spot was sent to the sensor
and the behavior of its response (i.e. the total integrated charge
of the sensor, pedestal subtracted)
was studied as a function
of the light intensity.

\begin{figure}[!h]
\centering
\includegraphics[width=0.34\linewidth,angle=90]{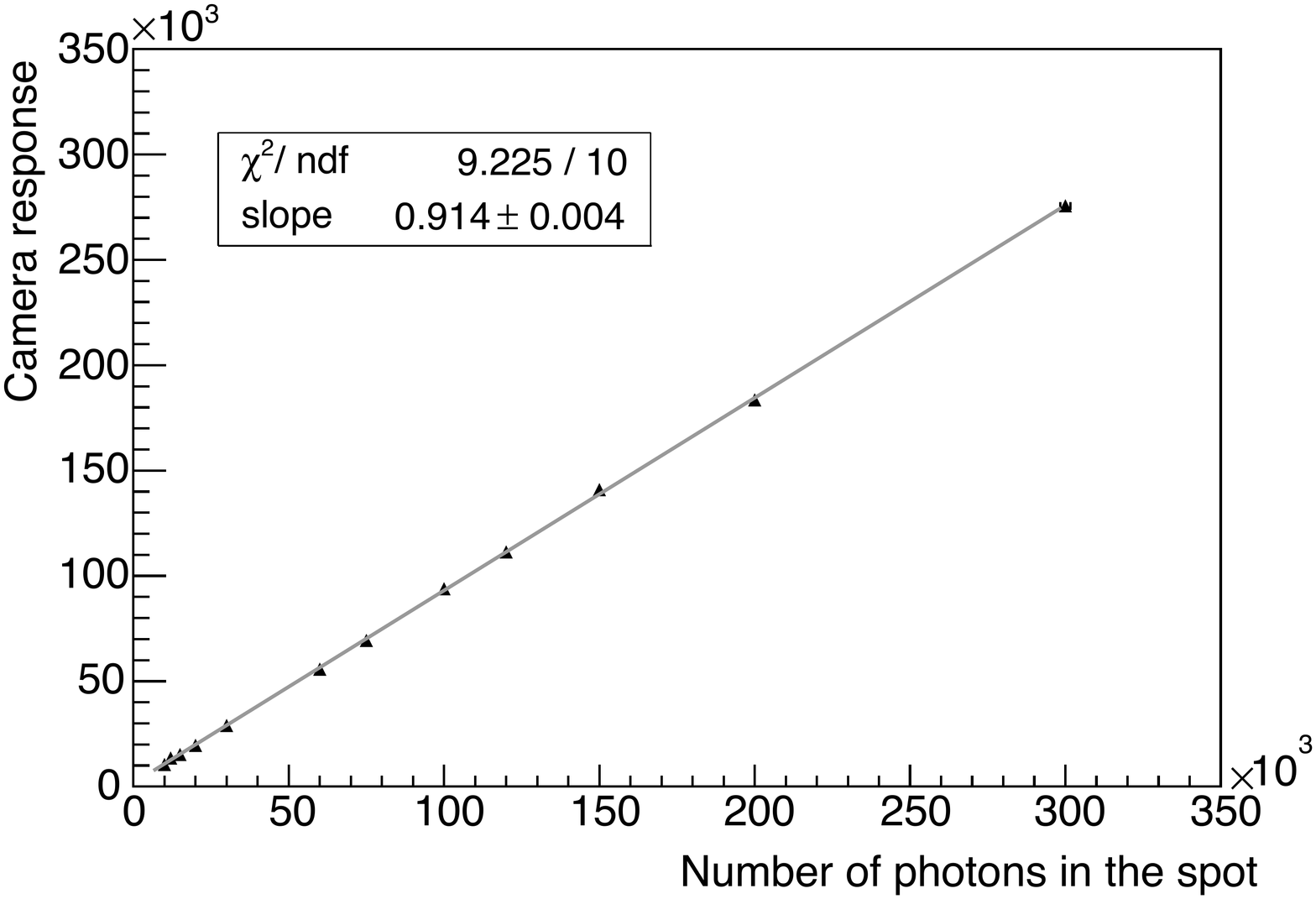}
\includegraphics[width=0.34\linewidth,angle=90]{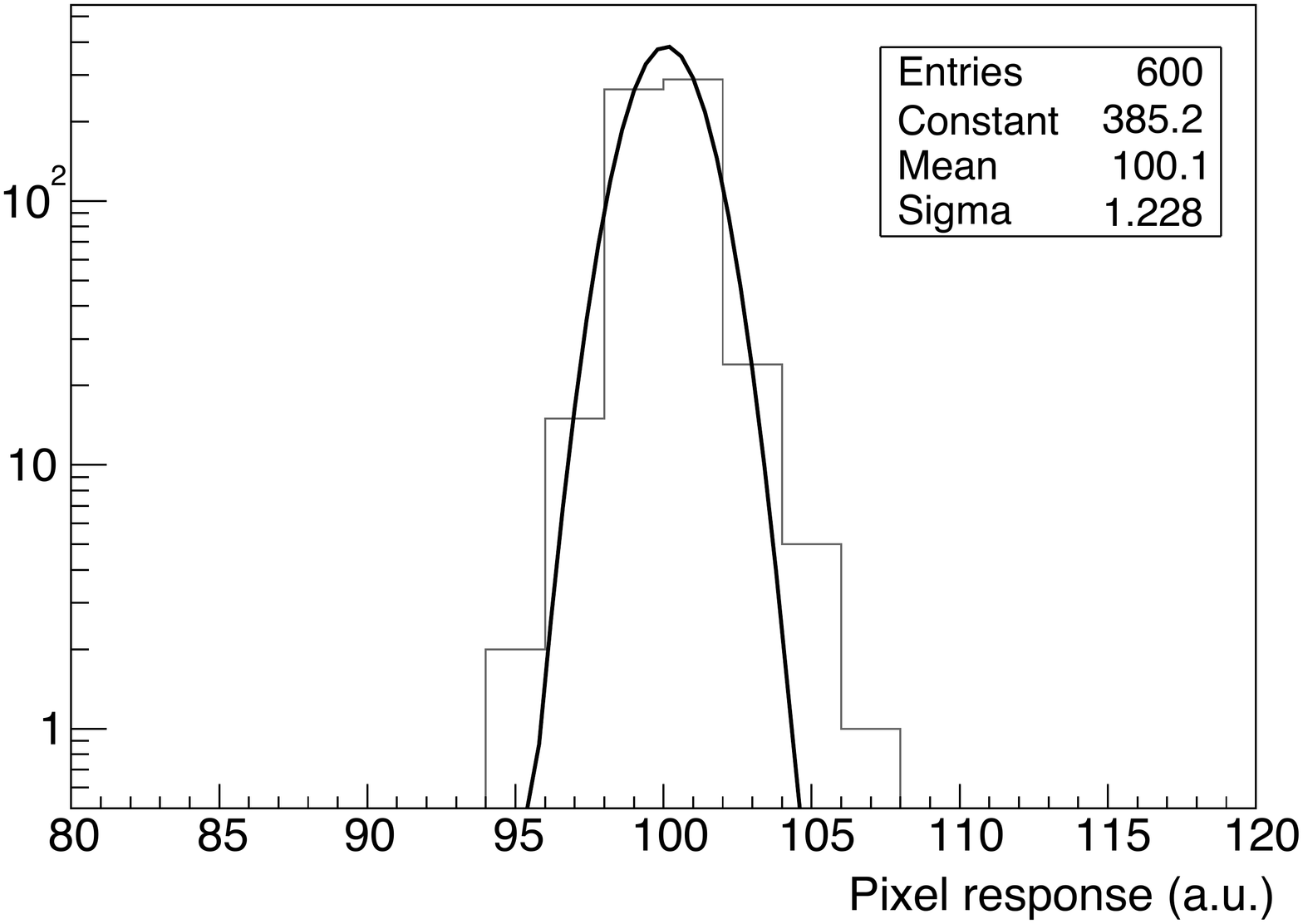}
\caption{
Left: behavior of the response of the camera as a function 
of the number of photons in the spot.
Right: distribution of the response of a not illuminated pixel.
}
\label{fig:camera}
\end{figure}

Figure \ref{fig:camera} shows on the left the results of this
study, with a superimposed linear fit.
The camera behaviour is well linear in the whole studied range
with a response of $0.91 \pm 0.01$ counts per photon.

In order to evaluate the noise level,
the response of a single pixel was acquired several times
while the camera was kept in the dark.
The distribution of the values obtained is shown on the right of
Fig. \ref{fig:camera} with a superimposed gaussian fit. 
Fluctuations of the pedestal are lower than 2 counts, i.e. lower
than two photons per pixel in good agreement with the expectations.

\subsection{Cosmic ray measurements}
\label{sec:mu}

The camera was equipped with
lens\footnote{17mm FL, f/0.95, Fast C-Mount. For more details visit
https://www.schneideroptics.com} 
providing an aperture of f/0.95 
excellent for working in low light conditions.
It was possible to acquire focused
images of the whole GEM area at a distance of 18 cm.
Each pixel looked at a surface of about
50$\mu$m $\times$ 50$\mu$m.

The experimental setup (Fig. \ref{fig:setup})
was rotated. The GEM and the gaps are put 
in vertical position. The aim is catching long tracks 
of muons crossing the drift gap almost parallel to the GEM plane. 
Thanks to the optical system it was possible to acquire
images of cosmic muons crossing the chamber and ionizing
the gas mixture within the drift gap.

\begin{figure}[!h]
\centering
\includegraphics[width=0.49\linewidth]{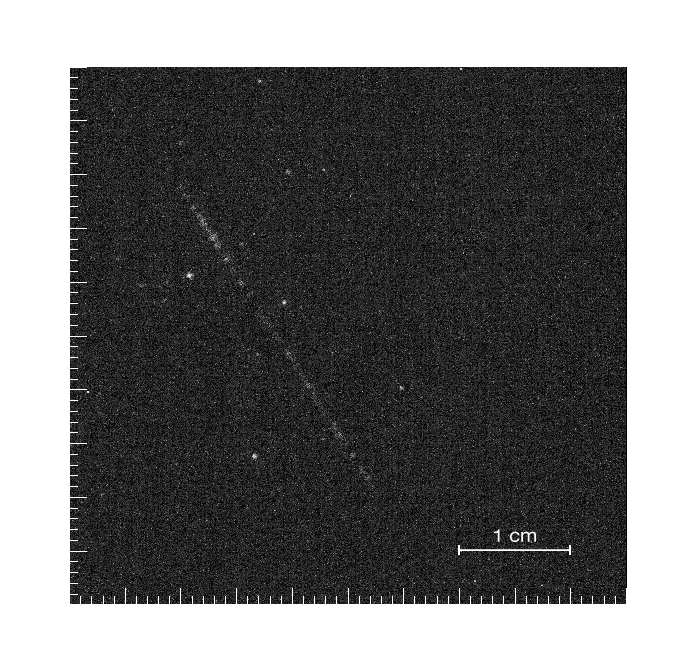}
\includegraphics[width=0.49\linewidth]{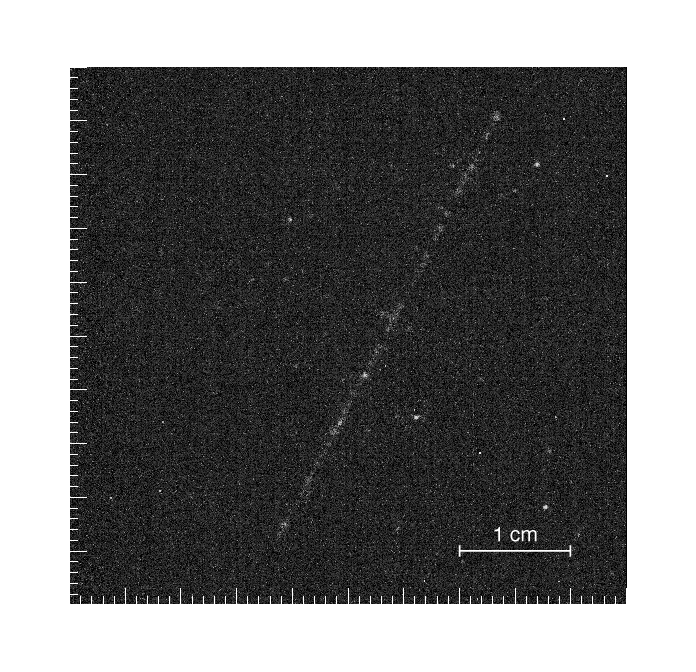}
\caption{Examples of cosmic track images acquired
by the CMOS based camera.} 
\label{fig:mu}
\end{figure}

Twenty tracks were collected in a five minute run.
Two examples of recorded images are displayed in Fig.~\ref{fig:mu}.
The pictures were taken with
an exposure of 100 ms. They show two tracks due to
particles crossing the detector. The one on the left is
about 29 mm long while the one on the right is about 42 mm. 

\begin{figure}[!h]
\centering
\includegraphics[width=0.49\linewidth]{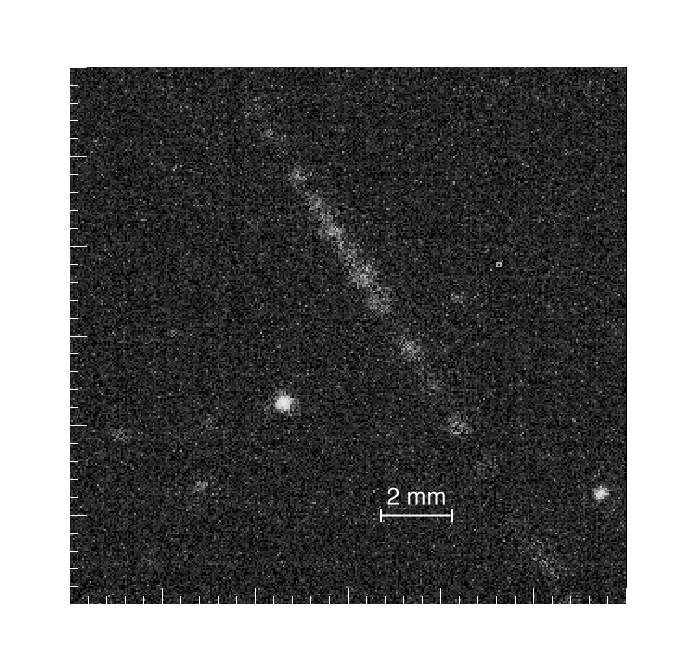}
\includegraphics[width=0.49\linewidth]{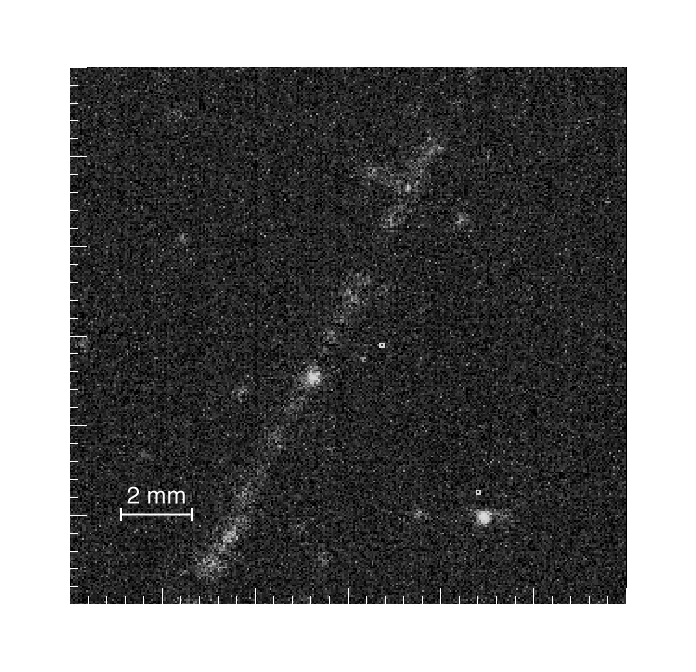}
\caption{Zoom on the track images acquired
by the CMOS based camera.} 
\label{fig:zoom}
\end{figure}

In Fig. \ref{fig:zoom} details of the track regions are shown.
The clusterization structure is well visible. 
Because of the electron diffusion in the gas, track images
are about 1 mm large.

\begin{figure}[!h]
\centering
\includegraphics[width=0.4\linewidth, angle=90]{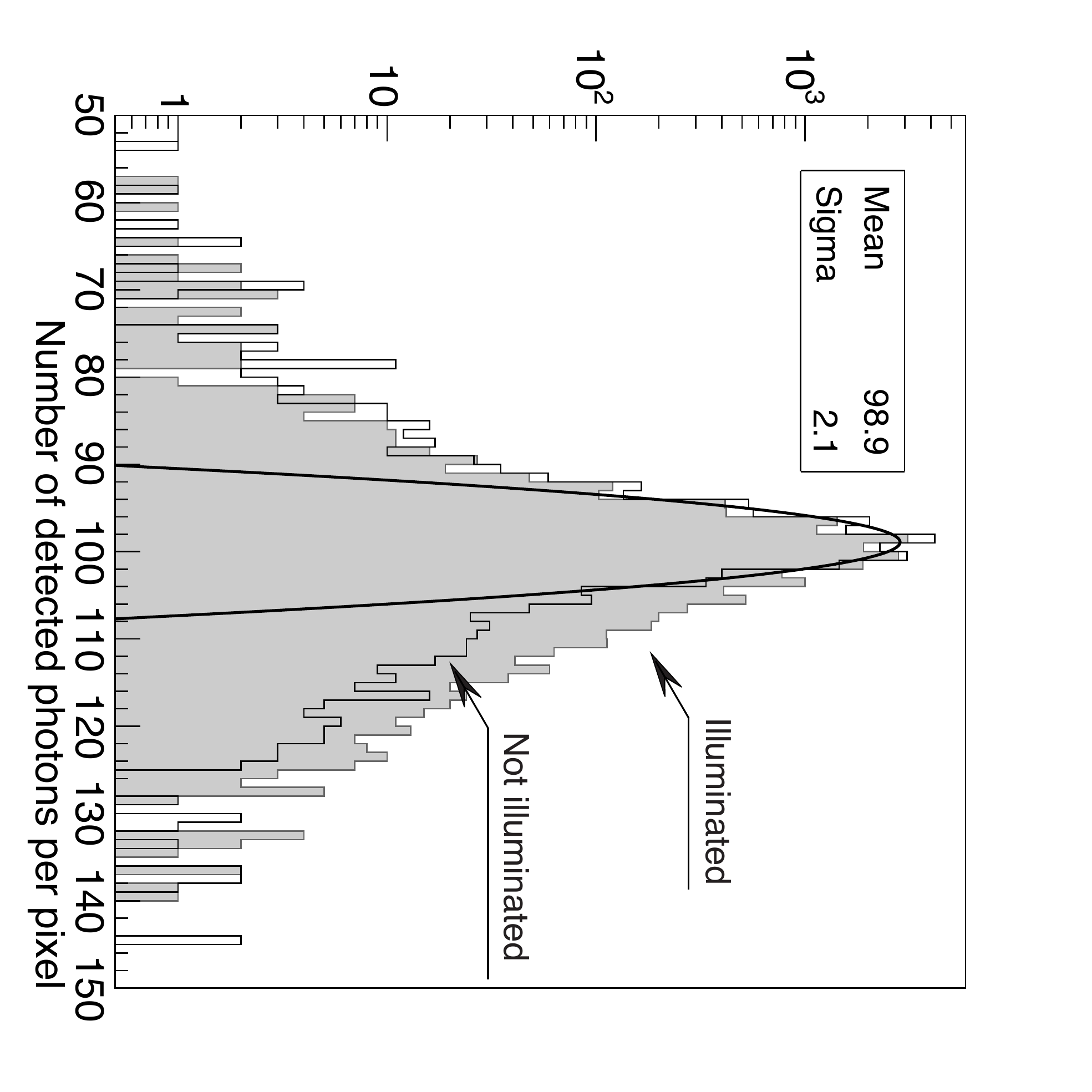}
\caption{Distributions of the pixel response inside and outside
the region illuminated by the track.}
\label{fig:qInOut}
\end{figure}

In order to get an idea about the amount of detected light
a simple analysis was performed. Figure \ref{fig:qInOut}
shows the distribution of the light detected by the pixels
in the track region and by pixels not illuminated by the track
for the event on the right in Fig. \ref{fig:mu}.
The superimposed gaussian fit demonstrates that the response of
not illuminated pixels
is $99 \pm 2$ counts in good agreement with the
result obtained in the dark condition
(Sect. \ref{sec:pre}).
The illuminated pixels, when properly accounted for the dark 
noise subtraction, are hence detecting up to 30 photons each.

The total amount of light integrated by the pixels in the track region
was thus evaluated by studing the recorded images.
It was measured that the linear light collection density is of about 
600$\pm$60 photons per track millimeter.
According to the simulation results (Sect. \ref{sec:gas})
7.7 primary electrons are produced per millimeter
by a cosmic ray. Therefore
about 80 photons are detected per primary electron.

\begin{figure}[!h]
\centering
\includegraphics[width=0.49\linewidth]{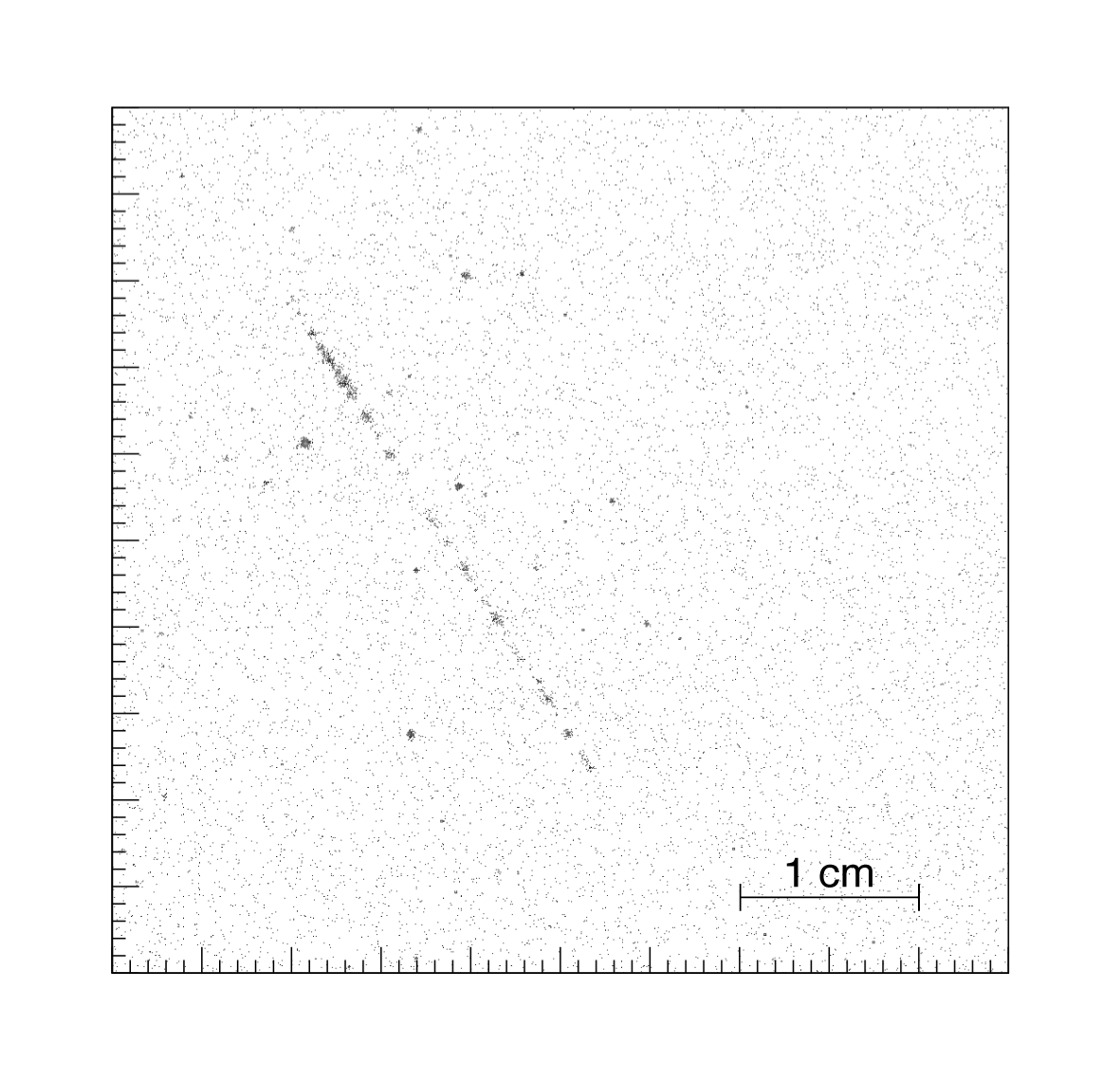}
\includegraphics[width=0.49\linewidth]{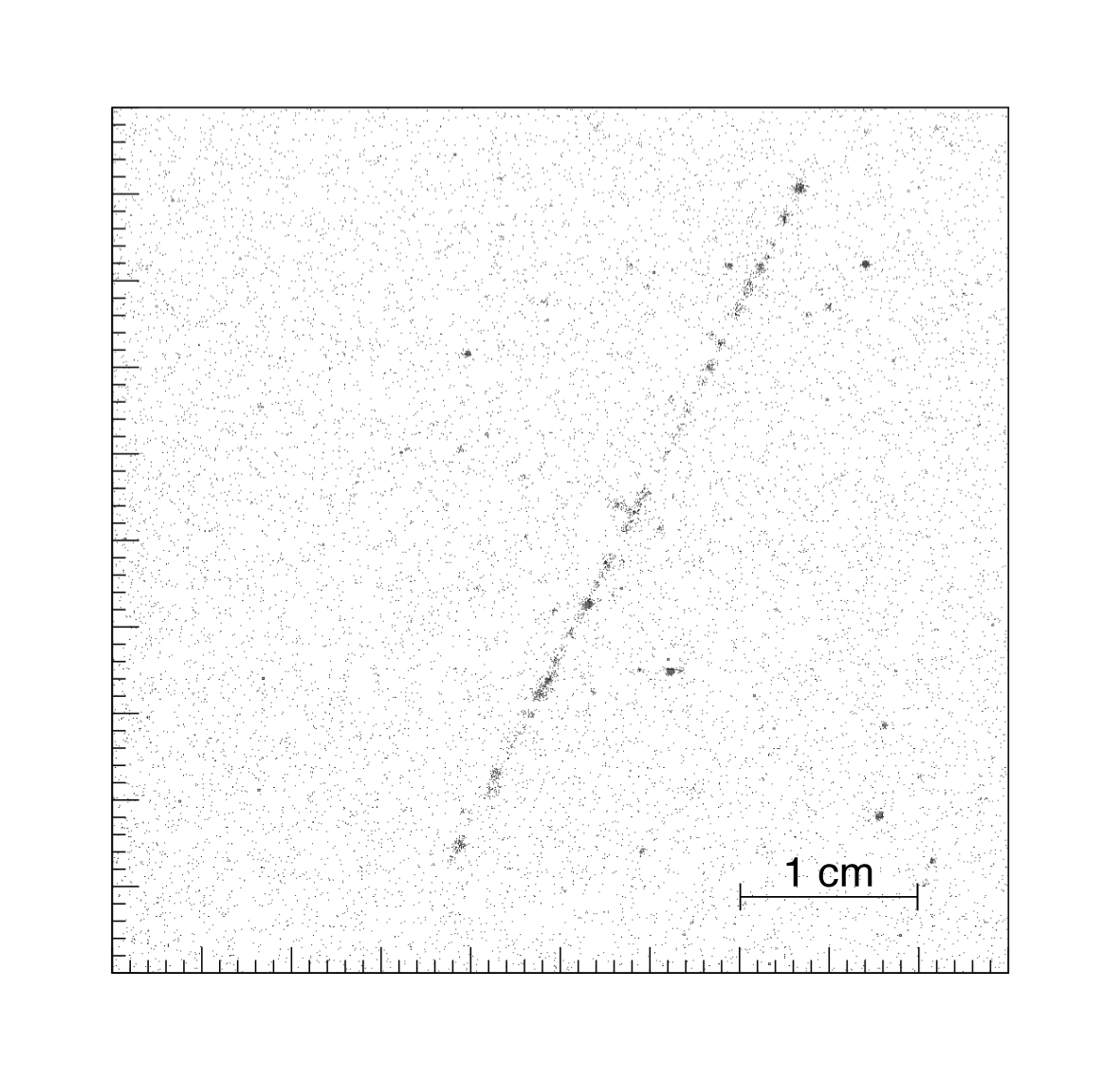}
\caption{Maps of the pixels with a response 
larger than 105 counts (i.e. three 
sigmas larger than the pedestals).}
\label{fig:mu3s}
\end{figure}

The maps of the pixels with a response three 
sigmas larger than the pedestal
(i.e. higher than 105 counts),
are shown Fig. \ref{fig:mu3s}. The tracks are well visible.
From the analysis of all the recorded muon tracks,
an amount of 40$\pm$5 pixels 
satisfying the above requirement
per track millimeter was measured.

\subsection{Electron measurements} 
 
\begin{figure}[!h]
\centering 
\includegraphics[width=0.45\linewidth]{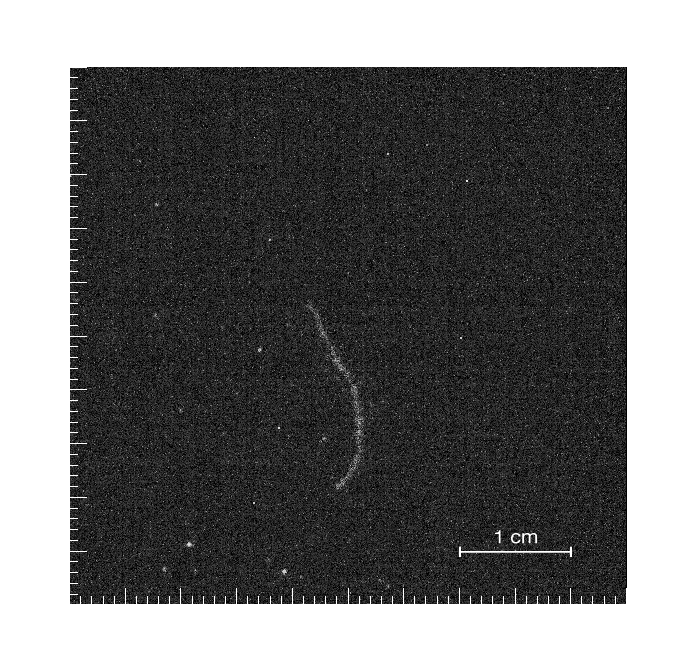} 
\includegraphics[width=0.45\linewidth]{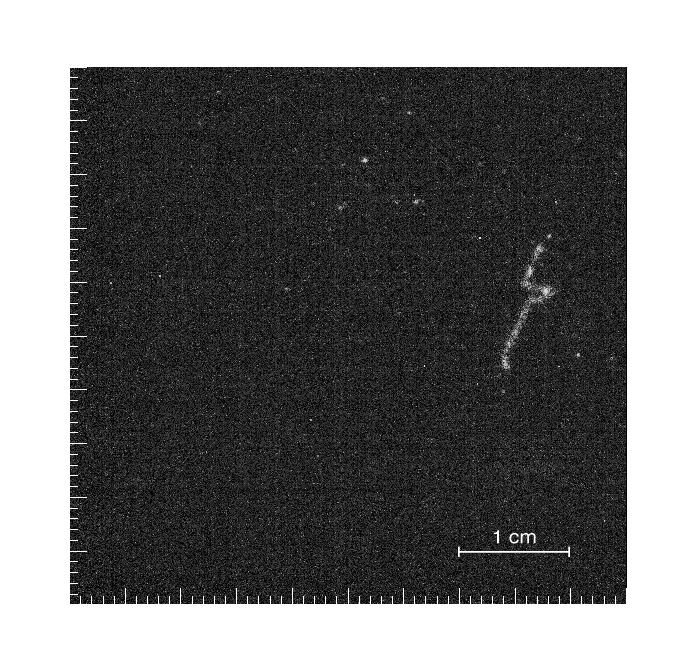} 
\caption{Examples of electron tracks.}  
\label{fig:ele} 
\end{figure}

\begin{figure}[!h]
\centering 
\includegraphics[width=0.45\linewidth]{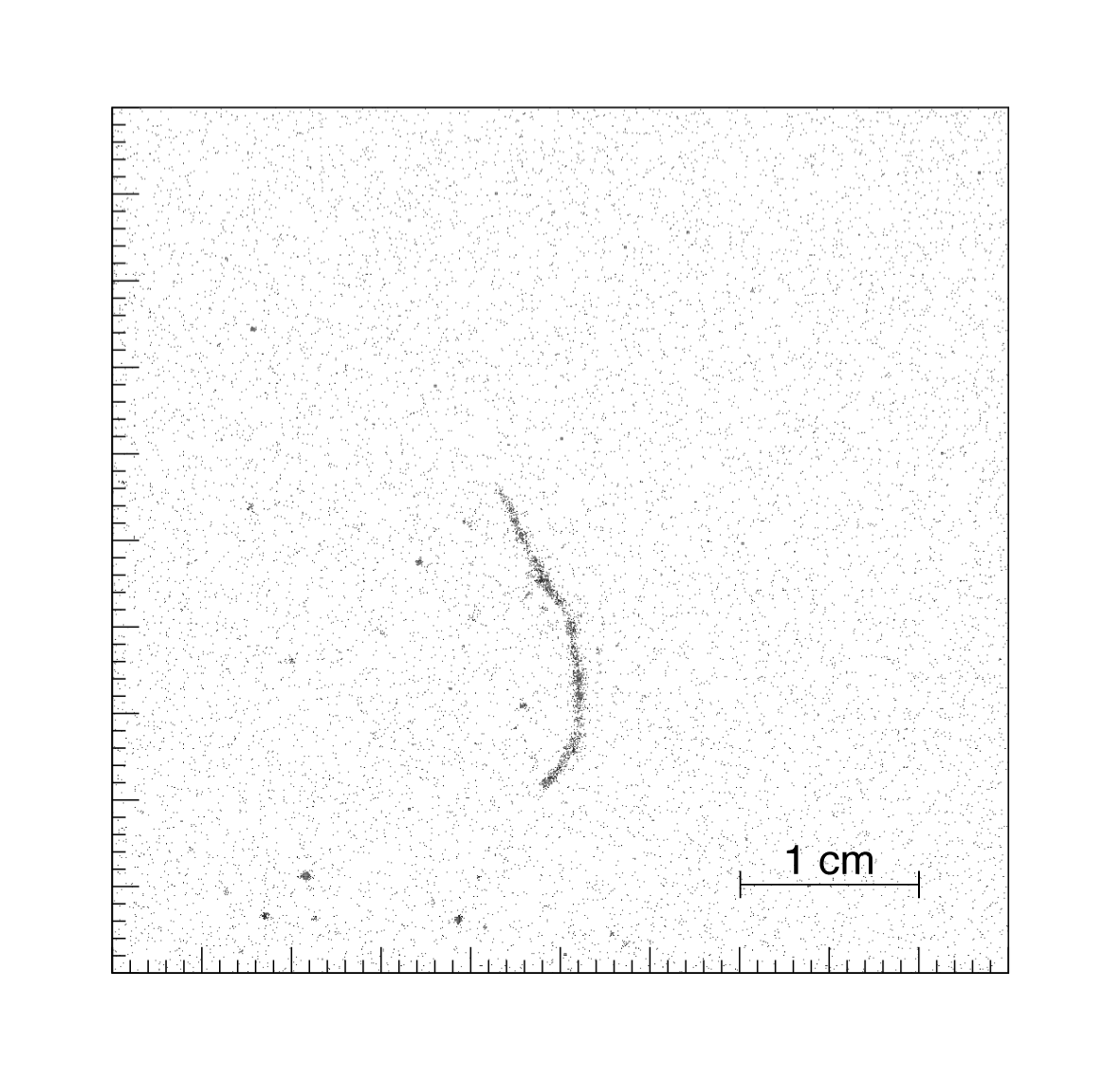} 
\includegraphics[width=0.45\linewidth]{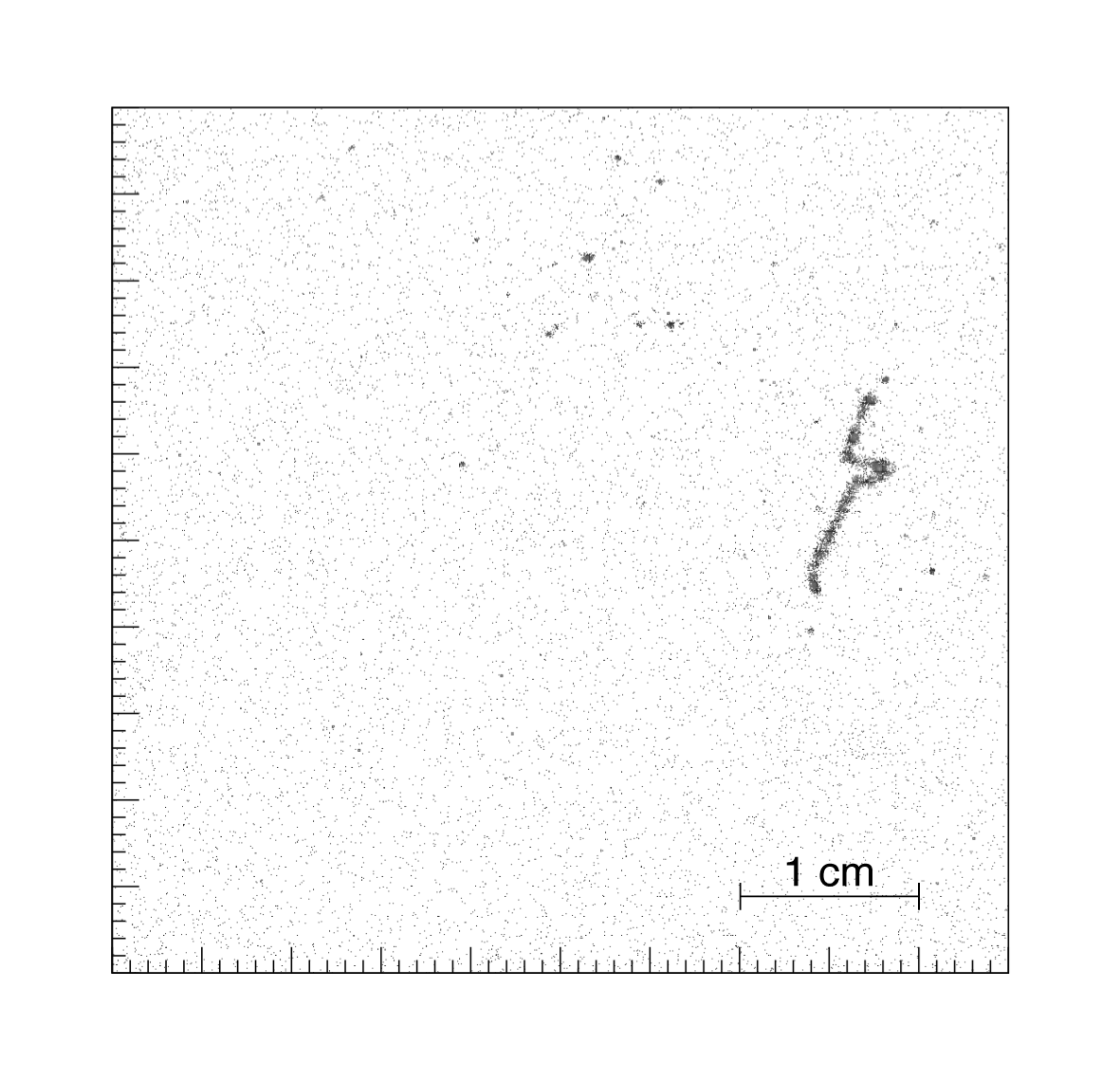} 
\caption{Maps of the pixels with a response 
larger than 105 counts (i.e. three 
sigmas larger than the pedestals).}
\label{fig:ele3s} 
\end{figure}

During the data taking, several images of short,  
intense and curved tracks were acquired. 
These tracks (as the ones shown in Fig. \ref{fig:ele}) 
are very likely due to ionizing electrons produced
by natural radioactivity and traveling within the  
drift gap. 
Tracks shown in Fig. \ref{fig:ele} are  long
approximatively 2.7 cm (left) and and 1.6 cm (right). 
As described in sect. \ref{sec:mu} 
the maps of pixels collecting an amount 
of light larger than three sigmas above the pedestal
were constructed.
The results are shown in Fig. \ref{fig:ele3s} 
 
As it is visible from the their intensities, 
the ionization density of these tracks is  
quite larger than the ones produced by cosmic ray muons.  
An amount of about 300 pixels  
three sigmas above the pedestal 
per track millimeter was measured. 

\section{Conclusion}
The light yield of a triple-GEM stack in a He/CF$_4$ gas mixture
was studied. After a suitable optimization of the electric fields,
a total amount of about 2500 photo-electrons was detected by means
of a PMT for a cosmic ray muon crossing a 3 mm wide gas gap.
Images of emitted light were recorded by a CMOS sensor. 
Thanks to its very high sensitivity
it has been possible to detect light on a large quantity of
pixels. 
Muon tracks, several centimeter longs,
where recorded.
For these minimum ionizing particles
average, in each track millimeter, 
more than 600 photons are detected 
making more than 30 pixels give
a response three sigmas above the pedestal.
Low energy electrons, due to natural radioactivity, 
were also detected. In these images, the amount of 
illuminated pixels is about ten times larger.

\section*{Acknowledgments}

The authors want to thank Marco Magi of the mechanics workshop of the
"Dipartimento di Scienze di Base e Applicate per l'Ingegneria"
for his fundamental support. 
This work took advantage of very interesting and useful 
discussions with the CERN-RD51 people.



\begin{thebibliography}{9}
\expandafter\ifx\csname url\endcsname\relax
  \def\url#1{\texttt{#1}}\fi
\expandafter\ifx\csname urlprefix\endcsname\relax\def\urlprefix{URL }\fi

\bibitem{Sauli:1997qp}
F.~Sauli, {GEM}: A new concept for electron amplification in gas detectors,
  Nucl. Instrum. Meth. A386 (1997) 531--534.

\bibitem{bib:fraga}
M. M. F. R. Fraga \textit{et al.},
"The GEM scintillation in He-CF$_4$, Ar-CF$_4$, Ar-TEA and Xe-TEA mixtures"
  Nucl. Instrum. Meth. A504 (2003) 88--92.

\bibitem{bib:tesi}
  D.~Pinci,
  ``A triple-GEM detector for the muon system of the LHCb experiment",
  CERN-THESIS-2006-070.

\bibitem{bib:garfield} 
  R.~Veenhof,
  ``Garfield, a drift chamber simulation program,''
  Conf.\ Proc.\ C {\bf 9306149}, 66 (1993).

\end{thebibliography}
\end{document}